\documentclass{article}
\usepackage{ijcai21}

\usepackage{times}
\usepackage{soul}
\usepackage{url}
\usepackage[hidelinks]{hyperref}
\usepackage[utf8]{inputenc}
\usepackage[small]{caption}
\usepackage{graphicx}
\usepackage{amsmath}
\usepackage{amsthm}
\usepackage{amsfonts}
\usepackage{booktabs}
\usepackage{algorithm}
\usepackage{algorithmic}
\usepackage{epsfig}
\usepackage{amssymb}
\usepackage{multirow}
\usepackage{float}
\usepackage{xspace}
\usepackage{color}
\usepackage[indention=10pt,position=top,margin=0pt,font=small,labelformat=parens,labelsep=space,skip=6pt,hypcap=false,labelfont=bf,list=true,textfont=sf]{subcaption}
\usepackage{pgfplots}
\usepackage{pifont}

\pgfplotsset{compat=1.16}

\title{\name: Backdoor Attack against Competitive Reinforcement Learning}

\author{
Lun Wang$^1$
\and
Zaynah Javed$^1$
\and
Xian Wu$^2$
\and
Wenbo Guo$^2$
\and
Xinyu Xing$^2$
\And 
Dawn Song$^1$
\affiliations
$^1$University of California, Berkeley\\
$^2$Pennsylvania State University
\emails
\{wanglun, zjaved\}@berkeley.edu,
xkw5132@psu.edu,
\{wzg13, xxing\}@ist.psu.edu,
dawnsong@cs.berkeley.edu
}

\begin{document}

\newcommand{\lun}[1]{\textcolor{blue}{Lun: #1}}
\newcommand{\zaynah}[1]{\textcolor{red}{Zaynah: #1}}
\newcommand{\name}{{\sc BackdooRL}\xspace}
\newtheorem{definition}{Definition}

\maketitle

\urlstyle{tt}

\begin{abstract}
Recent research has confirmed the feasibility of backdoor attacks in deep reinforcement learning (RL) systems.
However, the existing attacks require the ability to arbitrarily modify an agent's observation, constraining the application scope to simple RL systems such as Atari games.
In this paper, we migrate backdoor attacks to more complex RL systems involving multiple agents and explore the possibility of triggering the backdoor without directly manipulating the agent's observation.
As a proof of concept, we demonstrate that an adversary agent can trigger the backdoor of the victim agent with its own action in two-player competitive RL systems.
We prototype and evaluate \name in four competitive environments.
The results show that when the backdoor is activated, the winning rate of the victim drops by 17\% to 37\% compared to when not activated. 
The videos are hosted at https://github.com/wanglun1996/multi\_agent\_rl\_ backdoor\_videos.
\end{abstract}

\section{Introduction}

Backdoor attacks were originally proposed in \cite{gu2017badnets} as a new type of data poisoning attack for image classifiers.
Intuitively, backdoor attacks insert into a machine learning model some hidden functionality which can be activated by some specific triggers.
Recently, several works~\cite{yang2019design,kiourti2020trojdrl,wang2020stop} show that deep RL policies are also vulnerable to backdoor attacks.
However, these attacks have to manipulate the agent's observation directly and thus only work for RL tasks with totally tractable environments such as Atari games.
In this paper, we would like to migrate backdoor attacks to a wider range of RL tasks which involve complex interaction between the agents and the environment.
In particular, we focus on two-agent competitive reinforcement learning.

Competitive reinforcement learning has been applied in many safety-critical systems such as autonomous driving~\cite{shalev2016safe,dosovitskiy2017carla} and automated trading~\cite{dempster2006automated,noonan2017jpmorgan}. 
In these systems, the observation of an agent is the result of complex dynamics between the environment and the agents, and thus cannot be arbitrarily modified by the adversary.
Nevertheless, the adversary can still act as an agent and implicitly affect the observation by interacting with the environment.
For instance, an adversary can drive a car to change the camera input of another self-driving car but cannot make a landmark disappear.

In this paper, we propose \name, a backdoor attack targeted at two-player competitive reinforcement learning systems.
\name is different from traditional backdoor attacks in two perspectives.
First, the adversary agent has to lead the victim to take a series of wrong actions instead of only one to prevent it from winning.
Additionally, the adversary wants to exhibit the trigger action in as few steps as possible to avoid detection.
These contradictory requirements want the victim agent to memorize the transitory trigger to guide its long-term behavior. 
Second, the adversary has to trigger the victim's backdoor by interacting with the environment instead of directly modifying the input to the victim's policy.
The first challenge points us to policies based on sequential models (\emph{e.g.} RNN).
Sequential models can memorize the trigger and thus the adversary only needs to show the trigger in as little as one step.
However, the memory cannot last long so we would like the victim to be fast-failing after being triggered.
By adversarial policy training~\cite{huang2017adversarial,gleave} and reward manipulation, we manage to set up a unified framework for training a fast-failing agent.
To address the second challenge, we leverage imitation learning to embed both the normal and backdoor policies into the victim's policy.
To synthesize the trajectories for imitation learning, we mix the episodes of (1) an expert agent against a trigger-less agent, and (2) a fast-failing agent against the trigger agent.

We prototype \name in about 1700 lines of Python code and evaluate it in four different competitive environments.
The results show that when the trigger is not activated, the victim achieves comparable performance with policies without backdoors.
After the backdoor is triggered, the winning rate of the victim drops between 17\% to 37\%.
The videos of the simulated environments are hosted at https://github.com/wanglun1996/multi\_agent\_rl \_backdoor\_videos.
We also discuss possible defenses for \name and observe that fine-tuning can only weaken but not completely remove the backdoor functionality.

\vspace{2pt}
\paragraph{Contributions}
\begin{itemize}
    \item We propose \name, the first backdoor attack targeted at competitive reinforcement learning systems. The trigger is the action of another agent in the environment.
    \item We propose a unified method to design fast-failing agents for different environments.
    \item We prototype \name and evaluate it in four environments.
    The results validate the feasibility of backdoor attacks in competitive environments.
    \item We study possible defenses for \name. The results show that fine-tuning cannot completely remove the backdoor.
\end{itemize}
\section{Related Work}

The backdoor attack, where an adversary can alter an input to a neural network to trigger a hidden functionality, was discovered by~\cite{gu2017badnets} and has been implemented by~\cite{chen2017targeted,liu2017trojaning}~in image classifiers.
\cite{chen2020badnl} migrated backdoor attacks to NLP tasks.
Meanwhile, \cite{salem2020baaan} study backdoor attacks in generative adversarial networks (GANs).
Recently, there has been a line of work focusing on backdoor attacks under distributed learning or federated learning~\cite{bhagoji2019analyzing,xie2019dba,wang2020attack}.
Another spectrum of work studies backdoor attacks with a variety of properties such as clean-label backdoor~\cite{turner2018clean}, label-consistent backdoor~\cite{turner2019label}, triggerless backdoor~\cite{salem2020don} and hidden-trigger backdoor~\cite{saha2020hidden}.

Recent works point out that backdoor attacks also widely exist in RL tasks.
\cite{yang2019design} implant a backdoor in an agent walking a maze and successfully triggers the backdoor by changing the local landscape around the agent.
\cite{kiourti2020trojdrl} explore backdoor attacks in Atari games. 
These two attacks heavily rely on the manipulation of the environment and are limited to simple environments like Atari games. 
\cite{wang2020stop} go beyond the simple environments and manage to attack learning-based traffic congestion control systems.
In their attack, the adversary controls a malicious auto-vehicle and triggers the backdoor by acceleration or deceleration.
However, their attack depends on well-established principles of traffic physics.
Additionally, the victim (the congestion control system) does not interact with the vehicles so the attack can be viewed as a strengthened version of the previous attacks in which the adversary can only manipulate the observation in a limited way. 

A variety of detection methods and defenses have also been developed within the past few years.
\cite{chen2018detecting} propose to use activation clustering to detect anomaly patterns such as backdoors.
\cite{liu2018fine} find that fine-pruning can help remove existing backdoors in deep networks.
\cite{wang2019neural,guo2019tabor} detect backdoors by solving an optimization problem.
\cite{sun2019can} show that norm clipping and weak differential privacy can mitigate the attacks with only small performance overhead.
\cite{shan2020gotta} propose to use pre-inserted backdoors as honeypots to catch adversarial attacks.
Most defenses designed so far are for image classifiers.
\cite{zhu2020gangsweep} leverage GANs to sweep out neural backdoors.
We deem migrating the defenses to the competitive RL setting as an interesting future direction.

Other adversarial attacks on reinforcement learning have been explored in~\cite{huang2017adversarial,gleave}.
The attack in \cite{gleave} is modified and leveraged to train the fast-failing agent.
%
We refer interested readers to~\cite{lin2017tactics} for a systematic survey of adversarial attacks in RL.
\section{Background}

\subsection{Competitive Reinforcement Learning}

We model the competitive game as a two-player Markov process \cite{shapley1953stochastic}.

\begin{definition}[Two-player Markov decision process]
A two-player Markov decision process is comprised of a tuple $(\mathcal{S}, (\mathcal{A}_1, \mathcal{A}_2), T, (\mathcal{R}_1, \mathcal{R}_2))$.
$\mathcal{S}$ is the state space.
$\mathcal{A}_1$ and $\mathcal{A}_1$ denote the action spaces of the two agents.
$T: \mathcal{S}\times\mathcal{A}_1\times\mathcal{A}_2\rightarrow \mathcal{S}$ is the probabilistic transition function from $s\in \mathcal{S}$ to $s'\in \mathcal{S}$ conditioned on some action $(a_1, a_2)\in \mathcal{A}_1\times\mathcal{A}_2$.
$\mathcal{R}_1 (\mathcal{R}_2):\mathcal{S}\times\mathcal{A}_1\times\mathcal{A}_2\times \mathcal{S}\rightarrow \mathbb{R}$ is the reward functions for the two agents.
If $\mathcal{A}_1=\mathcal{A}_2$ and $\mathcal{R}_1=\mathcal{R}_2$, the process is called a symmetric MDP.
\end{definition}

If we assume one agent (\emph{e.g.} the second agent) follows a fixed stochastic policy $\pi_2$, then the two-player Markov decision process (MDP) reduces to a single-player MDP: $(\mathcal{S}, \mathcal{A}_1, T_1, \mathcal{R}_1')$.
The state and action space remain the same while the transition function and reward function is embedded with the second agent's policy $\pi_2$.
$$
T_1(s, a_1) = T_1(s, a_1, a_2)\text{ and }\mathcal{R}_1'(s, a_1, s')=\mathcal{R}_1(s, a_1, a_2, s')
$$
where $a_2\sim\pi_2(\cdot|s)$.
The agent's goal is to win a competitive game by maximizing the (discounted) accumulated reward 
\begin{equation}
\sum_{t=0}^\infty\gamma^t\mathcal{R}_1(s^{(t)}, a_1^{(t)}, s^{(t+1)})
\label{eq:reward}
\end{equation}
where $s^{(t+1)}\sim T_1(s^{(t)},a_1^{(t)})\text{ and }a_1\sim \pi_1(\cdot|s^{(t)})$.

\subsection{Imitation Learning}
\label{sec:bc}
Since its proposal, RL has become a powerful tool to solve many real-world control problems.
However, many systems either lack natural reward functions or have sparse rewards which thus requires people to manually design reward functions, which can be extremely challenging.
An alternative solution to RL is imitation learning.
In imitation learning, an expert demonstrates the desired behavior and the student directly imitates the expert's behavior.

Behavioral cloning is the simplest type of imitation learning, in which the student directly clones the expert's behavior using supervised learning.
Formally, we have trajectories $\{\tau_i|\tau_i\triangleq(s_i^{(0)}, a_i^{(0)}, s_i^{(1)}, a_i^{(1)}, s_i^{(2)}, \cdots)\}$ generated by an expert (\emph{e.g.} an experienced driver).
The learner's target is to mimic the expert's decisions as close as possible by finding a policy $\pi$ to minimize the loss function:
\begin{equation}
\sum_i\sum_t \ell(\pi(s_i^{(t)}), a_i^{(t)})    
\label{eq:loss}
\end{equation}
where $\ell(\cdot)$ is a distance function between the predicted action and the expert action such as MSE or cross-entropy.

\begin{figure}[t]
    \centering
    \includegraphics[width=0.45\textwidth]{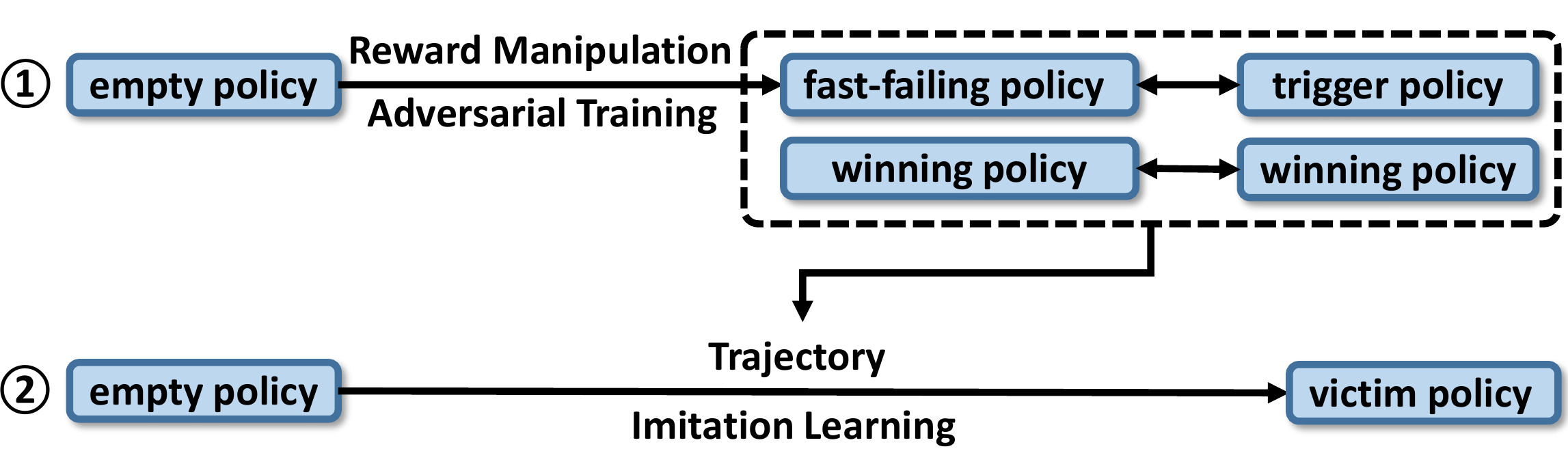}
    \caption{\name workflow. \ding{172} corresponds to the process of training the fast-failing agent while \ding{173} represents the imitation learning process. \ding{172} and \ding{173} are connected by the generation of trajectories.}
    \label{fig:workflow}
\end{figure}


\section{Methodology}

In this section, we formally present \name.
We first give an overview of \name and establish the notation system.
After that we introduce \name step by step including fast-failing agent training and backdoor functionality insertion.






\subsection{Overview \& Threat Model}

\begin{figure}[b]
    \centering
    \begin{subfigure}[b]{0.15\textwidth}
        \centering
        \includegraphics[width=\textwidth]{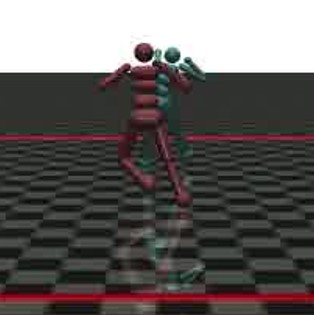}
        \caption{}
    \end{subfigure}\hfill
    \begin{subfigure}[b]{0.15\textwidth}
        \centering
        \includegraphics[width=\textwidth]{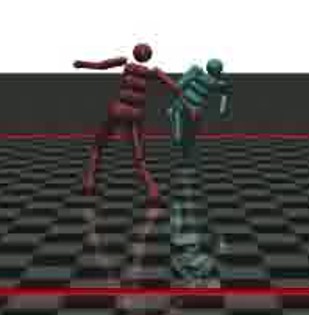}
        \caption{}
    \end{subfigure}\hfill
    \begin{subfigure}[b]{0.15\textwidth}
        \centering
        \includegraphics[width=\textwidth]{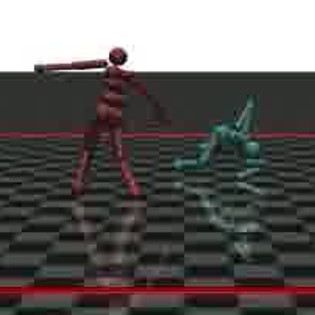}
        \caption{}
    \end{subfigure}\hfill
    

\caption{An illustration of \name in the Run to Goal and Sumo games. In (a), the adversary shows the trigger action; In (b), the backdoor in the victim is activated; In (c), the backdoor leads to the victim's failure.}
\label{fig:anim}
\end{figure}

\noindent\textbf{Overview.}
\name aims to insert a backdoor functionality into a victim policy denoted as $\pi_{\nu}$.
The backdoor can be activated by a series of trigger actions $(a_\text{trg}^{(0)}, a_\text{trg}^{(1)}, \cdots)$.
$\pi_{\nu}$ can be viewed as a mixture of two opposed policies.
When the backdoor is dormant, $\pi_{\nu}$ acts like a normal policy trying to win the game denoted as $\pi_\text{win}$.
When the backdoor is triggered, $\pi_\nu$ will try to fail as soon as possible like the fast-failing policy $\pi_\text{fail}$.
The attack workflow is illustrated in Figure~\ref{fig:anim}.

The complete pipeline is illustrated in Figure~\ref{fig:workflow}.
The first step is to generate the backdoor functionality $\pi_\text{fail}$.
Since an agent's memory does not last long, $\pi_\text{fail}$ is designed to fail in as few steps as possible.
To achieve this, we leverage adversarial training~\cite{gleave} and reward manipulation to train a fast-failing policy.
Concretely, we reduce the system to a single-player game by fixing the other agent's policy, and then train an adversarial policy by minimizing (instead of maximizing) the accumulated reward.
Since the original reward is intended to guide the agent to win, minimizing it leads the agent to fail fast.

In the second step, we synthesize a victim agent by hard-coding the following policy:
\begin{equation}
\pi_\text{hardcoded}(s)=\left\{
\begin{aligned}
\pi_\text{fail}(s) & \text{, if } \text{triggered} \\
\pi_\text{win}(s) & \text{, }o.w. \\
\end{aligned}
\right.
\label{eq:mixed}
\end{equation}
Note that although this policy acts perfectly as expected, it cannot be used in practice since it is easily detectable due to the explicit branch and an auxiliary boolean variable to denote ``triggered or not''.
Hence, we would like to train an LSTM-based policy to mimic the behavior of $\pi_\text{hardcoded}$.
To do so, we use behavioral cloning as discussed in Section \ref{sec:bc}.

\vspace{2pt}
\noindent\textbf{Threat Model.}
The studied RL systems involve two parties, each represented by an agent in the two-player game - a user and a malicious model developer.
The user would like to obtain an RL policy to perform her task.
She either outsources the training job to the malicious developer or downloads a pre-trained policy from the developer.
Therefore, the developer can arbitrarily deviate from the normal training procedure to implant a backdoor in the output policy.
The user, on the other hand, is able to distinguish a malicious policy by examining anomaly program structures but cannot find a backdoor hidden in the parameters of a neural network. 

\subsection{Train the Fast-failing Agent}

In \name, we would like the trigger and the backdoor to be as stealthy as possible to avoid possible detection.
Thus, the adversary agent wants to exhibit the trigger behavior for as few steps as possible.
This requires the victim agent (\emph{e.g.} following a LSTM-based policy) to remember the trigger and execute the backdoor functionality even after the trigger action disappears in the other agent.
Although LSTM can memorize longer than RNN~\cite{hochreiter1997lstm}, the memory length is still limited to hundreds of steps and gradually decays with time.
Hence, we want the backdoor functionality to lead the victim to fail as fast as possible.

\begin{figure}[t]
    \centering
    \begin{subfigure}[b]{0.15\textwidth}
        \centering
        \includegraphics[width=\textwidth]{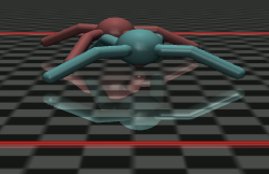}
        \caption{ }
        \label{fig:rtgants_illustration}
    \end{subfigure}
    \hfill
    \begin{subfigure}[b]{0.15\textwidth}
        \centering
        \includegraphics[width=\textwidth]{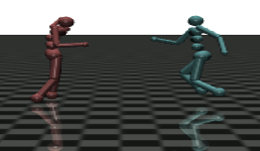}
        \caption{ }
        \label{fig:ysnp_illustration}
    \end{subfigure}
    \hfill
    \begin{subfigure}[b]{0.15\textwidth}
        \centering
        \includegraphics[width=\textwidth]{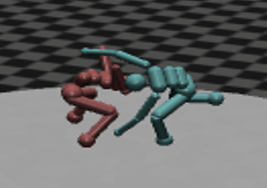}
        \caption{ }
        \label{fig:shumans_illustration}
    \end{subfigure}
    
\caption{Illustrations for environments from~\protect\cite{bansal2017emergent}. (a) Run to Goal (ants); (b) Run to Goal (humans) / You Shall Not Pass (humans); (c) Sumo (humans).}
\label{fig:illustration}
\end{figure}

The simplest backdoor is to stay still and wait for the other agent to beat it.
However, in many tasks, doing nothing might result in a slow failure.
For instance, in the Run to Goal game (Ants) in Figure ~\ref{fig:rtgants_illustration}, staying still will block the other agent's path and cost the other agent many steps to get around the obstacle.
An ideal fast-failing agent should first give way to the adversary and then stay still.
However, designing such ad hoc failing behaviors is time-consuming and requires expertise about the environment.

We design a unified method to obtain the fast-failing agent, inspired by the adversarial attack in~\cite{gleave}.
The attack in~\cite{gleave} embeds a fixed policy in the two-agent environment, maximizes the accumulated reward defined in~\ref{eq:reward} and thus obtains an adversarial policy which can quickly beat the embedded agent.
We take a similar approach to obtaining the fast-failing agent by minimizing the accumulated reward (or equivalently, maximizing the opposite of Equation ~\ref{eq:reward}).
To accelerate the failure, we introduce a constant penalty reward  $c$ $(c<0)$ for each time-step.
Thus, our method can be viewed as maximizing the following reward:
$$
\sum_{t=0}^\infty\gamma^t(c-\mathcal{R}_1(s^{(t)}, a_1^{(t)}, s^{(t+1)})).
$$

\subsection{Mimic the Hardcoded Agent}
\label{sec:mimic}

We want the final victim policy to preserve the following two properties:
\begin{itemize}
        \item {\bf Effectiveness}: The backdoor agent should fail with high probability when the adversary shows the trigger.
        \item {\bf Stealthiness}: (1) The behavior of the trigger agent should be normal when facing normal agents. (2) The trigger itself should be as stealthy as possible to avoid human auditing and detection of the attack.
         
\end{itemize}

The first step is to generate the trajectories for imitation learning.
We first hard-encode a victim policy as Equation~\ref{eq:mixed}, and an adversary agent:
$$
\pi_\text{adv}(s)=\left\{
\begin{aligned}
&a_\text{trg}^{(0)}, \text{cnt++}, \text{flag=T} \text{ \textbf{if} flag=F}\text{, w.p. } p_\text{trg}\\
&a_\text{trg}^{(cnt)}, \text{cnt++} \text{ \textbf{if} flag=T, cnt}\leq|(a_\text{trg}^{(0)}, a_\text{trg}^{(1)}, \cdots)| \\
&\pi_\text{win}(s), \text{cnt}=0, \text{flag=F} \text{ \textbf{if} cnt}>|(a_\text{trg}^{(0)}, a_\text{trg}^{(1)}, \cdots)| \\
&\pi_\text{win}(s) \text{ \textbf{if} flag=F, w.p. } 1-p_\text{trg} \\
\end{aligned}
\right.
$$
Concretely, the adversary agent plays the trigger action with probability $p_\text{trg}$ and otherwise follows the winning policy $\pi_\text{win}$.
We play the victim agent against the adversary agent to generate the trajectories $\{(s_i^{(0)}, a_i^{(0)}, s_i^{(1)}, a_i^{(1)}, s_i^{(2)}, \cdots)\}$.
Then we use behavior cloning to burn the mixed policy into one deep policy by minimizing~\ref{eq:loss}.

\noindent\textbf{Remark 1.} The generated trajectory is a mixture of two different types of episodes.
If the adversary does not show the trigger, the generated episode is normal.
By learning from these episodes, the victim will act normally when facing normal agents.
Otherwise, the episode encodes a series of behaviors: (1) see the trigger; (2) activate backdoor; (3) fail.
By imitating these episodes, the victim can learn the backdoor functionality.

Note that the hard-coded victim agent has access to the boolean indicator flag of the adversary agent during the simulation but the indicator is not included in the states in the trajectories.
By doing so, we hope the victim policy learns to capture the trigger action by observing the effect of the trigger action on the observation.

\noindent\textbf{Remark 2.} To satisfy stealthiness, we want the length of the trigger behavior to be as short as possible.
Hence, the victim policy should be able to memorize the state of being triggered as long as possible even after the trigger disappears.
As a result, we target LSTM-based policies.

\pgfplotsset{every axis/.append style={
                    label style={font=\Huge},
                    tick label style={font=\Huge}  
                    }}

\begin{figure*}[t]
    \centering
    
    \begin{subfigure}[b]{0.2\textwidth}
        \centering
        \resizebox{\textwidth}{!}{\begin{tikzpicture}
\begin{axis}[
  set layers,
  grid=major,
  ytick align=outside, ytick pos=left,
  xtick align=outside, xtick pos=left,
  xlabel=\# Epochs,
  ylabel={Policy Loss},
  legend style={fill opacity=0.6, draw opacity=1, text opacity=1, at={(1.5,0.4)}, anchor=north, nodes={scale=0.75, transform shape}},
  ]

\addplot+[
  dashed, line width=1.6pt, mark options={scale=0.75},
  smooth, 
  error bars/.cd, 
    y fixed,
    y dir=both, 
    y explicit
] table [x=epoch, y=loss, y error=error, col sep=comma] {data/adv/rtgants.txt};

\end{axis}
\end{tikzpicture}}
        \caption{Run to Goal (ants).}
        \label{fig:rtgants}
    \end{subfigure}
    \begin{subfigure}[b]{0.2\textwidth}
        \centering
        \resizebox{\textwidth}{!}{\begin{tikzpicture}
\begin{axis}[
  set layers,
  grid=major,
  ytick align=outside, ytick pos=left,
  xtick align=outside, xtick pos=left,
  xlabel=\# Epochs,
  ylabel={Policy Loss},
  legend style={fill opacity=0.6, draw opacity=1, text opacity=1, at={(1.5,0.4)}, anchor=north, nodes={scale=0.75, transform shape}},
  ]

\addplot+[
  dashed, line width=1.6pt, mark options={scale=0.75},
  smooth, 
  error bars/.cd, 
    y fixed,
    y dir=both, 
    y explicit
] table [x=epoch, y=loss, y error=error, col sep=comma] {data/adv/rtghumans.txt};

\end{axis}
\end{tikzpicture}}
        \caption{Run to Goal (humans).}
        \label{fig:rtghumans}
    \end{subfigure}
    \begin{subfigure}[b]{0.2\textwidth}
        \centering
        \resizebox{\textwidth}{!}{\begin{tikzpicture}
\begin{axis}[
  set layers,
  grid=major,
  ytick align=outside, ytick pos=left,
  xtick align=outside, xtick pos=left,
  xlabel=\# Epochs,
  ylabel={Policy Loss},
  legend style={fill opacity=0.6, draw opacity=1, text opacity=1, at={(1.5,0.4)}, anchor=north, nodes={scale=0.75, transform shape}},
  ]

\addplot+[
  dashed, line width=1.6pt, mark options={scale=0.75},
  smooth, 
  error bars/.cd, 
    y fixed,
    y dir=both, 
    y explicit
] table [x=epoch, y=loss, y error=error, col sep=comma] {data/adv/ysnphumans.txt};

\end{axis}
\end{tikzpicture}}
        \caption{You Shall Not Pass (humans).}
        \label{fig:ysnphumans}
    \end{subfigure}
    \begin{subfigure}[b]{0.2\textwidth}
        \centering
        \resizebox{\textwidth}{!}{\begin{tikzpicture}
\begin{axis}[
  set layers,
  grid=major,
  ytick align=outside, ytick pos=left,
  xtick align=outside, xtick pos=left,
  xlabel=\# Epochs,
  ylabel={Policy Loss},
  legend style={fill opacity=0.6, draw opacity=1, text opacity=1, at={(1.5,0.4)}, anchor=north, nodes={scale=0.75, transform shape}},
  ]

\addplot+[
  dashed, line width=1.6pt, mark options={scale=0.75},
  smooth, 
  error bars/.cd, 
    y fixed,
    y dir=both, 
    y explicit
] table [x=epoch, y=loss, y error=error, col sep=comma] {data/adv/shumans.txt};

\end{axis}
\end{tikzpicture}}
        \caption{Sumo (humans).}
        \label{fig:shumans}
    \end{subfigure}
    \hfill

    \caption{Policy Loss vs. Epoch for adversarial training.}
    \label{fig:advloss}
    \vspace{-10pt}
\end{figure*}
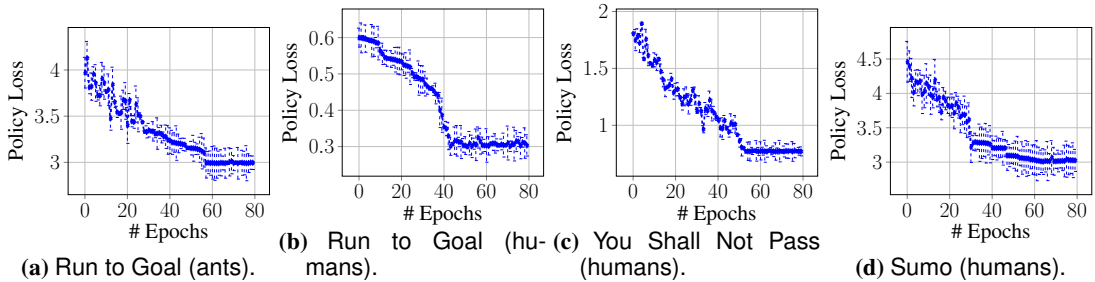
\section{Evaluation}

In this section, we empirically validate the feasibility of \name in four competitive RL environments.
Specifically, we would like to answer the following questions:
(1) Does the fast-failing agent obtained by adversarial training fail faster than dummy agents which passively wait for failure?
(2) Is the backdoor in the victim activated when facing the trigger action?
(3) How does the victim agent perform without observing the trigger?
(4) How do the hyper-parameters influence the attack?

\subsection{Experiment Setup}

To answer the questions, we evaluate \name in four different environments~\cite{bansal2017emergent}.
The agents we trained to use in the environments are of two possible types:
    \begin{itemize}
        \item {\bf Humanoid}: Bipedal, upright body with 17 joints.
        \item {\bf Ant}: Quadrupedal body with 8 joints.
    \end{itemize}
The competitive game environments (Figure~\ref{fig:illustration}) used to test our agents are:
    \begin{itemize}
        \item {\bf Run to Goal}: Two agents are set onto a flat space with two finish lines, facing one another. The agent that reaches its finish line first wins. 
        \item {\bf You Shall Not Pass}: Two agents are set onto a flat space, facing one another. One agent is the blocker while the other one is the runner who wants to run to the finish line that is behind the blocker. The runner wins if it makes to the finish line. The blocker wins if the runner does not make it. No ties are possible in this game.
        \item {\bf Sumo}: Two agents are set onto a circular arena space, facing one another. The agent will win if it has touched the other agent and is standing after the other agent has fallen. In the event that the opponent falls before being touched, the game results in a tie. Therefore, failure of the agents to make any physical contact will result in a tie.
    \end{itemize}

We implement \name in about 1700 lines of Python code. 
For adversarial training, we leverage an implementation of Proximal Policy Optimization (PPO) from Stable Baselines~\cite{stable-baselines}.
For the victim policy, we use a two-layer LSTM with MLP feature extraction as implemented in~\cite{stable-baselines}.

\subsection{Evaluation Results}
In this section, we present the evaluation results and answer question (1), (2), (3) and (4) empirically.

\vspace{2pt}
\noindent\textbf{Fast-failing Agent.}
To answer question (1), we first evaluate the process of fast-failing agent training.
The loss curves for the adversarial training are shown in Figure~\ref{fig:advloss} and the error bars are obtained across three independent trainings.
The adversarial training typically needs 40 to 60 epochs to converge.

After convergence, we play the fast-failing agent against a normal agent and record the steps needed before failure as shown in Table~\ref{tab:ffa}.
\begin{table}[b]
\centering
\small
\begin{tabular}{|l|c|c|}
\hline
Environment & Fast-failing & Dummy \\
\hline
Run to Goal (Ants) & \textbf{223.48} & 265.79 \\ \hline
Run to Goal (Humans) & \textbf{53.56} & 109.87 \\ \hline
You Shall Not Pass (Humans) & \textbf{62.96} & 71.68 \\ \hline
Sumo (Humans) & \textbf{40.23} & 60.87 \\ \hline
\end{tabular}
\vspace{-5pt}
\caption{Average steps to fail (fast-failing agent vs. dummy agent).}
\label{tab:ffa}
\end{table}
We also record the steps before failure for a dummy agent which takes no action.
By comparing the number of steps before failure, we find that the fast-failing agent saves 12.2\% to 51.3\% steps, relieving the pressure to memorize the trigger.

To further verify the effectiveness of the fast-failing agent, we evaluate the length of the episodes of the victim agent.
In a random sample of 500 games, the length of the episodes generally fell within a range from 40-100 timesteps, with most episodes being roughly 60-70 timesteps.
%
%
%
This confirms that the generated policy is able to end in a failure in a quick manner.

\noindent\textbf{Behavioral Cloning.}
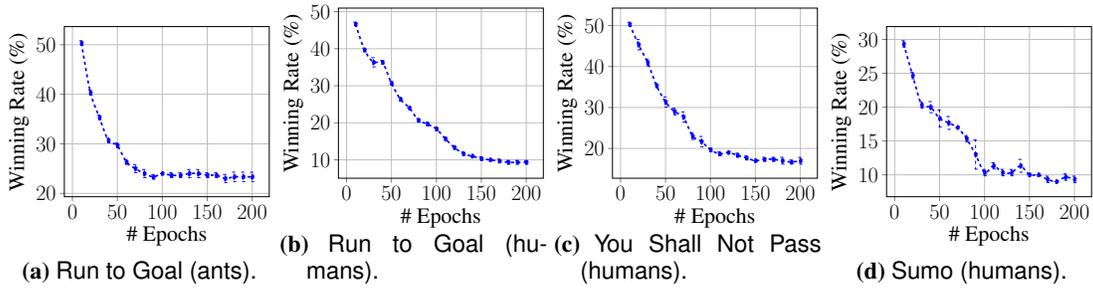
\begin{figure*}[t]
    \centering
    
    \begin{subfigure}[b]{0.2\textwidth}
        \centering
        \resizebox{\textwidth}{!}{\begin{tikzpicture}
\begin{axis}[
  set layers,
  grid=major,
  ytick align=outside, ytick pos=left,
  xtick align=outside, xtick pos=left,
  xlabel=\# Epochs,
  ylabel={Winning Rate (\%)},
  legend style={fill opacity=0.6, draw opacity=1, text opacity=1, at={(1.5,0.4)}, anchor=north, nodes={scale=0.75, transform shape}},
  ]

\addplot+[
  dashed, line width=1.6pt, mark options={scale=0.75},
  smooth, 
  error bars/.cd, 
    y fixed,
    y dir=both, 
    y explicit
] table [x expr=(\coordindex+1)*10, y=rate, y error=error, col sep=comma] {data/mimic/rtgants.txt};

\end{axis}
\end{tikzpicture}}
        \caption{Run to Goal (ants).}
        \label{fig:rtgants_imi}
    \end{subfigure}
    \begin{subfigure}[b]{0.2\textwidth}
        \centering
        \resizebox{\textwidth}{!}{\begin{tikzpicture}
\begin{axis}[
  set layers,
  grid=major,
  ytick align=outside, ytick pos=left,
  xtick align=outside, xtick pos=left,
  xlabel=\# Epochs,
  ylabel={Winning Rate (\%)},
  legend style={fill opacity=0.6, draw opacity=1, text opacity=1, at={(1.5,0.4)}, anchor=north, nodes={scale=0.75, transform shape}},
  ]

\addplot+[
  dashed, line width=1.6pt, mark options={scale=0.75},
  smooth, 
  error bars/.cd, 
    y fixed,
    y dir=both, 
    y explicit
] table [x expr=(\coordindex+1)*10, y=rate, y error=error, col sep=comma] {data/mimic/rtghumans.txt};

\end{axis}
\end{tikzpicture}}
        \caption{Run to Goal (humans).}
        \label{fig:rtghumans_imi}
    \end{subfigure}
    \begin{subfigure}[b]{0.2\textwidth}
        \centering
        \resizebox{\textwidth}{!}{\begin{tikzpicture}
\begin{axis}[
  set layers,
  grid=major,
  ytick align=outside, ytick pos=left,
  xtick align=outside, xtick pos=left,
  xlabel=\# Epochs,
  ylabel={Winning Rate (\%)},
  legend style={fill opacity=0.6, draw opacity=1, text opacity=1, at={(1.5,0.4)}, anchor=north, nodes={scale=0.75, transform shape}},
  ]

\addplot+[
  dashed, line width=1.6pt, mark options={scale=0.75},
  smooth, 
  error bars/.cd, 
    y fixed,
    y dir=both, 
    y explicit
] table [x expr=(\coordindex+1)*10, y=rate, y error=error, col sep=comma] {data/mimic/ysnphumans.txt};

\end{axis}
\end{tikzpicture}}
        \caption{You Shall Not Pass (humans).}
        \label{fig:ysnphumans_imi}
    \end{subfigure}
    \begin{subfigure}[b]{0.2\textwidth}
        \centering
        \resizebox{\textwidth}{!}{\begin{tikzpicture}
\begin{axis}[
  set layers,
  grid=major,
  ytick align=outside, ytick pos=left,
  xtick align=outside, xtick pos=left,
  xlabel=\# Epochs,
  ylabel={Winning Rate (\%)},
  legend style={fill opacity=0.6, draw opacity=1, text opacity=1, at={(1.5,0.4)}, anchor=north, nodes={scale=0.75, transform shape}},
  ]

\addplot+[
  dashed, line width=1.6pt, mark options={scale=0.75},
  smooth, 
  error bars/.cd, 
    y fixed,
    y dir=both, 
    y explicit
] table [x expr=(\coordindex+1)*10, y=rate, y error=error, col sep=comma] {data/mimic/shumans.txt};

\end{axis}
\end{tikzpicture}}
        \caption{Sumo (humans).}
        \label{fig:shumans_imi}
    \end{subfigure}
    \hfill

    \caption{Winning Rate vs. Epoch for imitation learning.}
    \label{fig:mimicrate}
\end{figure*}
\begin{table*}[t]
\centering
\small
\begin{tabular}{|l|c|c|c|c|c|c|c|c|c|}
\hline
\multicolumn{1}{|c|}{\multirow{2}{*}{Environment}} & \multicolumn{3}{c|}{Triggered} & \multicolumn{3}{c|}{Not Triggered} & \multicolumn{3}{c|}{Benign Baseline} \\ \cline{2-10} 
\multicolumn{1}{|c|}{} & \multicolumn{1}{c|}{Failing} & Tie & Winning & Failing & Tie & Winning & Failing & Tie & Winning\\ \hline
Run To Goal (Ants) & 73.8\% & 2.4\% & 23.8\% & 45.0\% & 5.1\% & 49.9\% & 46.0\% & 3.0\% & 51.0\%\\ \hline
Run To Goal (Humans) & 20.8\% & 69.3\% & 9.9\% & 52.2\% & 0.7\% & 47.1\% & 51.2\% & 1.4\% & 47.4\%\\ \hline
You Shall Not Pass (Humans) & 83.0\% & 0.0\% & 17.0\% & 50.1\% & 0.0\% & 49.9\% & 50.5\% & 0.0\% & 49.5\%\\ \hline
Sumo (Humans) & 34.4\% & 54.7\% & 10.9\% & 29.7\% & 42.2\% & 28.1\% & 30.1\% & 34.4\% & 35.5\%\\ \hline
\end{tabular}
\caption{The failing/tie/winning rate of the victim agent when the backdoor is triggered (or not). Benign baselines are measured on two normal agents.}
\label{tab:results}
\vspace{-10pt}
\end{table*}
To answer questions (2) and (3), we use behavior cloning to mix the fast-failing agent and the winning agent in one LSTM.
The winning rate vs. epoch curves are shown in Figure~\ref{fig:mimicrate}.
For each environment, 1000 games were simulated to determine the winning rates and the error bars are obtained across three independent training.
The agent stably learns the backdoor functionality after around 150 epochs, mimicking the synthesized trajectory.

We then show the victim trigger behavior to test the effectiveness of the backdoor as shown in Table~\ref{tab:results} to answer (2).
We also run the victim agent against normal agents to answer (3).
The results show that the failing rate of the victim agent increases by 4\% to 33\% and the winning rate drops by 17\% to 37\% due to the existence of ties in the game.
An increase in the tie rate can either be a positive or a negative result.
Here the increase is negative since the winning rate drops and the failing rate increases.
When running without the trigger, the victim agent achieves comparable performance to the normal agent.

Across all experiments, the trigger was indiscernible to the human eye.
We choose staying still as the trigger for all four environments.
The reason is two-fold and contradictory: (1) staying still is more stealthy than exhibiting abnormal actions; (2) the trigger action cannot be overly similar to actions performed by a normal agent.
Hence, staying still is a perfect balance between too ``obvious'' and too ``stealthy''.
In all of the environments, with the exception of the Sumo (Humans) environment, the trigger behavior was placed at the beginning of the game. 
For the Sumo (Humans) environment, the trigger has to be delayed because the winning condition for the Sumo game requires that the agents make contact with one another or else the game is considered to be a tie. 
We find that delaying the game around 60 timesteps into the game gives the best results.
This provides a noticeable improvement from the baseline winning rate of the victim.

\noindent\textbf{Trigger Length.}
To answer question (4), we evaluated the influence of trigger length on the attack. 
The results are shown in Table~\ref{tab:trigger}.
In general, the longer the trigger, the higher the win rate.
However, a long trigger is not as stealthy to the eye.

\begin{table}[htbp]
    \small
    \centering
    \begin{tabular}{|c|c|c|c|}
    \hline
    Environment & Trigger Length & Winning Rate & Tie Rate \\
    \hline
    \multirow{6}{*}{RTG(A)} & 5 & 40.9\% & 3.0\%\\
    \cline{2-4}
    & 10 & 31.7\% & 3.1\% \\
    \cline{2-4}
    & 30 & 16.5\% & 5.2\% \\
    \cline{2-4}
    & 50 & 17.4\% & 4.6\% \\
    \cline{2-4}
    & 100 & 5.0\% & 0.7\% \\
    \hline
    \multirow{6}{*}{RTG(H)} & 5 & 11.2\% & 84.8\%\\
    \cline{2-4}
    & 10 & 14.9\% & 84.6\% \\
    \cline{2-4}
    & 30 & 0.0\% & 85.5\% \\
    \cline{2-4}
    & 50 & 0.0\% & 95.0\%\\
    \cline{2-4}
    & 100 & 0.0\% & 90.6\% \\
    \hline
    \multirow{6}{*}{YSNP(H)} & 5 & 27.8\% & 0.0\% \\
    \cline{2-4}
    & 10 & 19.9\% & 0.0\% \\
    \cline{2-4}
    & 30 & 0.0\% & 0.0\% \\
    \cline{2-4}
    & 50 & 0.0\% & 0.0\% \\
    \cline{2-4}
    & 100 & 0.0\% & 0.0\% \\
    \hline
    \multirow{6}{*}{S(H)} & 5 & 15.9\% & 81.3\%\\
    \cline{2-4}
    & 10 & 8.7\% & 86.8\% \\
    \cline{2-4}
    & 30 & 10.0\% & 57.9\% \\
    \cline{2-4}
    & 50 & 9.1\% & 55.7 \\
    \cline{2-4}
    & 100 & 8.7\% & 54.5\% \\
    \hline
    \end{tabular}
    \caption{Winning/Tie rate of the victim agent with different trigger lengths (1000 episodes). The names of environments are abbreviated.}
    \label{tab:trigger}
\vspace{-10pt}
\end{table}

\noindent\textbf{Other Trigger Patterns.}
To answer question (4), we evaluate two other trigger patterns as shown in Table~\ref{tab:pattern}.
The results show that different patterns will also work but the attack performance and stealthiness might differ.

\begin{table}[htbp]
    \small
    \centering
    \begin{tabular}{|c|c|c|c|}
    \hline
    Environment & Trigger Pattern & Winning Rate & Tie Rate \\
    \hline
    RTG(A) & \ding{172} & 22.9\% & 6.4\%\\
    \hline
    \multirow{2}{*}{RTG(H)} & \ding{172} & 2.9\% & 80.0\%\\
    \cline{2-4}
    & \ding{173} & 4.0\% & 74.2\% \\
    \hline
    \multirow{2}{*}{YSNP(H)} & \ding{172} & 21.8\% & 0.0\% \\
    \cline{2-4}
    & \ding{173} & 0.0\% & 0.0\% \\
    \hline
    \multirow{2}{*}{S(H)} & \ding{172} & 6.0\% & 85.5\%\\
    \cline{2-4}
    & \ding{173} & 3.7\% & 89.0\% \\
    \hline
    \end{tabular}
    \caption{Winning/Tie rate of the victim agent with different trigger patterns (1000 episodes). \ding{172}: Shift to the left; \ding{173}: Left arm bent.}
    \label{tab:pattern}
\vspace{-10pt}
\end{table}

\noindent\textbf{Poisoning Rate.}
To answer question (4), we study the effect of poisoning rate, shown in Table~\ref{tab:poison}.
The higher the poisoning rate, the more harmful but the less stealthy the attack is.

\begin{table}[htbp]
    \small
    \centering
    \begin{tabular}{|c|c|c|c|}
    \hline
    Poisoning Rate & Winning Rate & Tie Rate & Failing Rate \\
    \hline
    10\% & 44.2\% & 3.0\% & 52.5\%\\
    \hline
    20\% & 40.7\% & 1.5\% & 57.8\%\\
    \hline
    30\% & 32.6\% & 1.3\% & 66.1\%\\
    \hline
    40\% & 26.0\% & 3.4\% & 70.6\%\\
    \hline
    \end{tabular}
    \caption{Winning/Tie rate of the victim agent with different poisoning rate (Run To goal (Ants), 1000 episodes).}
    \label{tab:poison}
\vspace{-10pt}
\end{table}

\section{Defenses}
\label{sec:defence}

In this section, we discuss possible defenses for \name.
One possible defense is to fine tune (or un-learn) the victim network by retraining with additional normal episodes~\cite{liu2018fine}
%
To test out this potential defense, we take the victim agents we have generated and generate 1000 normal episodes and train the victim agent for 200 epochs. 
We test the agents again in the same four environments for 1000 games each.

\begin{table}[ht]
\centering
\small
\begin{tabular}{|l|c|c|}
\hline
Environment/Failing Rate               & \name & Fine-tuned \\ \hline
Run to Goal (Ants)          & 23.8\% & \textbf{39.0\%}       \\ \hline
Run to Goal (Humans)        & \textbf{9.9\%} & 5.0\%       \\ \hline
You Shall Not Pass & 17.0\% & \textbf{23.8\%}      \\ \hline
Sumo            & 10.9\% & \textbf{22.5\%}      \\ \hline
\end{tabular}
\caption{Winning rate before/after fine-tuning when facing the trigger. Bolded numbers are the higher winning rates.}
\label{tab:fine}
\vspace{-5pt}
\end{table}
%
As shown in Figure~\ref{tab:fine}, for 3 out of 4 environments, the winning rates of the victim agents improve from 6.8\% to 15.2\%.

The winning rate of Run to Goal (Humans) instead drops by 4.9\%.
We conjecture that the fine-tuning process leads to over-fitting which degrades the performance.

Additionally, we notice that even fine-tuning for more epochs cannot further improve the winning rate.
Concretely, we fine-tune for another 500 epochs and notice that the winning rate stays roughly the same within 1\%. 
We conjecture that fine-tuning can only enhance the winning policy but cannot make the agent totally forget the failing policy since the trigger action is seldom seen in a normal trajectory.
As a conclusion, fine-tuning is only effective in a extremely limited sense.
It requires information about the trigger behavior and might even lead to performance drop. 

Researchers have developed various defences for backdoor attacks in image classifiers~\cite{chen2018detecting,liu2018fine,wang2019neural,guo2019tabor,sun2019can,shan2020gotta,zhu2020gangsweep} and we deem it as an important future direction to migrate these defenses to \name and validate their effectiveness.
\vspace{-5pt}
\section{Conclusion}
We have concluded that it is in fact possible to insert a backdoor into deep reinforcement learning, allowing an adversary to use some malicious input actions to force its opponent to fail. 
This model of attack is of critical significance since it universally exists in many safety-critical machine learning and robotic systems, such as self-driving cars and trading systems.
%



\vspace{-10pt}
\section*{Acknowledgments}

This material is in part based upon work supported by DARPA contract \#N66001-15-C-4066, the Center for Long-Term Cybersecurity, and Berkeley Deep Drive. 
Any opinions, findings, conclusions, or recommendations expressed in this material are those of the authors, and do not necessarily reflect the views of the sponsors.

\footnotesize
\bibliographystyle{named}
{\linespread{0.5}\selectfont\bibliography{ref}}









\end{document}